\documentclass[a4paper, twoside, 12pt]{article}
\usepackage[utf8]{inputenc}
\usepackage[T1]{fontenc}
\usepackage{graphicx}
\usepackage{tabularx,longtable}
\usepackage{hyperref}
\usepackage{caption}
\usepackage{url}
\usepackage{bibnames}

\usepackage[left=2.5cm, right=2.5cm, top=2.5cm, bottom=2.5cm, bindingoffset=1.5cm, head=15pt]{geometry} 
\usepackage{setspace}
\usepackage{fancyhdr,amsmath,amssymb}
\usepackage{algpseudocode}
\usepackage{algorithm}

\let\bold\textbf

\title{A practical analysis of ROP attacks}

\newcommand{\thesisauthor}{Ayush Bansal} 
\newcommand{\studentID}{160177} 
\newcommand{\thesistype}{UGP Report} 
\newcommand{\supervisor}{Dr. Debadatta Mishra}

\begin{document}

\makeatletter
\begin{titlepage}
    \begin{center}
        \vspace*{1cm}

        \Large
        \textbf{\@title}

        \vspace{1.5cm}
        
        \thesistype{}
        
        \vspace{1cm}

        \begin{figure}[htbp]
             \centering
             \includegraphics[width=.5\linewidth]{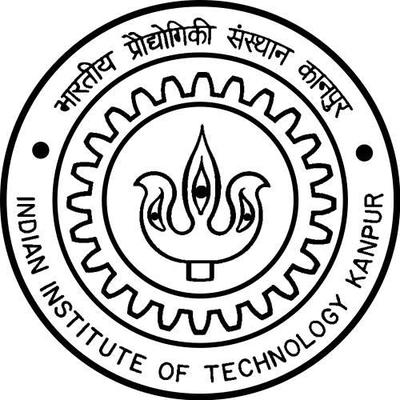}
        \end{figure}

        \vspace{1cm}

        \large
        \textbf{Author}: \thesisauthor{} (Student ID: \studentID{})\\
        \large
        \textbf{Advisor}: \supervisor{}\\

        \vspace{1cm}
        \large
        Department of Computer Science and Engineering\\
        Indian Institute of Technology, Kanpur\\

        \vspace{1cm}
        April 27, 2019

    \end{center}
\end{titlepage}
\makeatother


\clearpage
\thispagestyle{empty}
\section*{Abstract}

Control Flow Hijacking attacks have posed a serious threat to the security of applications for a long time where an attacker can damage the control Flow Integrity of the program and execute arbitrary code. These attacks can be performed by injecting code in the program’s memory or reusing already existing code in the program (also known as Code-Reuse Attacks). Code-Reuse Attacks in the form of Return-into-libc Attacks or Return-Oriented Programming Attacks are said to be Turing Complete, providing a guarantee that there will always exist code segments (also called ROP gadgets) within a binary allowing an attacker to perform any kind of function by building a suitable ROP chain (chain of ROP gadgets).

Our goal is to study different techniques of performing ROP Attacks and find the difficulties encountered to perform such attacks. For this purpose, we have designed an automated tool which works on 64-bit systems and generates a ROP chain from ROP gadgets to execute arbitrary system calls.

\clearpage
\pagenumbering{Roman}
\tableofcontents
\clearpage

\pagenumbering{arabic}


\clearpage

\section{Introduction}
\label{sec:Intro}

Security community has faced the problem of malicious code execution for a long time. The attacker gaining access can steal sensitive data or hijack an application's execution. Although, the attacker needs to divert the application's control flow to execute a code of his choosing. If the attacker is successful, the program can be used for arbitrary purposes, such as attacking other systems or applications.

A problem which persists while preventing these attacks is, one cannot accurately predict whether a particular execution will be benign or not. For a long time, people have concentrated on preventing the introduction and execution of new malicious code. The techniques used for preventing these attacks roughly fall into two categories, attempting to guarantee the control flow integrity in existing programs (e.g., type-safe languages, stack cookies, XFI \cite{XFI}), and attempting to isolate malicious code that has been introduced into the system (e.g. W$\oplus$X, memory tainting). The execution of injected code can be prevented using W$\oplus$X (Writable/Executable) or DEP (Data Execution Prevention) \cite{DEP}

Our discussion will be on a category of control flow hijacking attacks which do not require any code injection by the attacker, yet can induce arbitrary behaviour in the targeted system, namely Code-Reuse attacks. In said attacks, fragments of existing code present in the program's address space (\textit{gadgets}) are linked together to perform malicious execution. Since the code being executed is in the executable (code) region of the program's address space, DEP or other techniques will no longer be viable. Aforementioned gadgets can be put together in an order resulting in consecutive execution to perform an arbitrary function.

Many techniques have been introduced that allow putting gadgets together resulting in consecutive execution. The most widespread one is called Return-Oriented Programming (ROP) \cite{ROP-1}, \cite{ROP-2}. Another one is called Jump-Oriented Programming (JOP) \cite{JOP-1}, \cite{JOP-2}, \cite{JOP-3}, which not used popularly as it is difficult to facilitate a deeper comprehension of code-reuse-attacks using this technique. ROP makes use of code snippets which end in a \texttt{ret} instruction which is just like a \texttt{pop rip} instruction.

Many tools are available for finding ROP gadgets in an executable file, our tool retrieves gadgets using \textit{ROPgadget} tool by Jonathan Salwan \cite{ROPgadget}. This tool retrieves ROP gadgets upto an arbitrary maximum depth from an executable. The gadgets thus obtained can be chained together to perform arbitrary computation.

Research done in this area during the last decade and more has suggested that ROP is Turing Complete \cite{ROP-1}, \cite{ROP-2}, i.e. it is in theory possible to use ROP gadgets to construct any arbitrary program. In fact, ROP has been shown to be Turing Complete on several platforms \cite{ROP-3}, \cite{ROP-1}, \cite{ROP-2}. This statement implies that any shellcode an attacker wants to execute in a regular code injection attack, he/she can also create using ROP, thus bypassing DEP and making ROP extremely dangerous.

Creating a program using ROP is, however, complex, cumbersome, and requires a lot of manual labour and attention to detail. In regard of this problem, we have designed a tool to analyze the binary input file and create a suitable ROP chain, thus automating the process of program creation using ROP.

During the course of our discussion, we pick a system architecture. We work with binaries on \texttt{x86\_64} architecture - specifically, the Standard C Library on Linux 64-bit systems. The reasons behind this choice are:

\begin{itemize}
  \setlength\itemsep{0em}
  \item It is one of the most widely used architecture in current computer systems.
  \item The \texttt{x86\_64} architecture uses registers for parameters during system call and function call sequences, this makes finding a suitable ROP gadget chain more challenging.
\end{itemize}

We infer from our experience that any sufficiently large body of executable code on any architecture and operating system will feature sequences that allow the construction of similar gadgets.

\noindent Our goals regarding this analysis are:
\begin{enumerate}
  \setlength\itemsep{0em}
  \item Identify the issues, encountered while creating a program using ROP, which make it complex and cumbersome to work with. Later work on solutions of these issues.
  \item Since ROP attacks are said to be Turing Complete, we want to automate the process of generating a program using ROP.
  \item Verify the Turing Completeness property of Return-Oriented programming.
\end{enumerate}

\noindent Our contributions:
\begin{itemize}
    \setlength\itemsep{0em}
    \item Designed an efficient scheme for transforming gadgets into a sequence of simple equations.
    \item Designed an algorithm to output a permutation of ROP gadgets executing an arbitrary program.
    \item Extended the functionality of the \textit{ROPgadget} tool to produce a ROP chain along with the list of gadgets.
\end{itemize}

\clearpage
\section{Background}
\label{sec:Background}

In this section, our discussion will be on ways of performing a control flow hijacking attack, then we will move on to implications of Return-Oriented programming on DEP and present day systems. Lastly, we will discuss the \texttt{x86\_64} architecture.

\subsection{Performing a Control Flow Hijacking Attack}
Consider an attacker who has discovered a vulnerability in an application and wants to exploit it. Exploitation, here, means subverting the application's control flow so that the application performs attacker-directed actions with its permissions or credentials. One of the most famous such vulnerability class is the \textit{stack bufferoverflow} \cite{Aleph}, other famous classes of such type are buffer overflows on the heap i.e. \textit{heap overflow} \cite{Heap-1}, \cite{Heap-2}, \cite{Heap-3}, \textit{integer overflows} \cite{int-1}, \cite{int-2}, \cite{int-3}, and \textit{format string vulnerabilities} \cite{str-1}, \cite{str-2}.

To complete the objective, the attacker must (1) subvert the program's control flow from its original course, and (2) redirect the program's execution to the code of his/her choosing. To complete task (1), attacker can perform the traditional stack-smashing attack (e.g. stack buffer overflow) and overwrite the return address of the function on the stack, so it points to code of his/her choosing rather than to the caller function. Here, other techniques may also be used such as \textit{frame-pointer overwriting} \cite{frame}. The attacker can complete task (2) by injecting code into the program's memory; the modified return address should be pointed to this code. Usually the injected code is \textit{shellcode}, whether or not it spawns a shell.

We will restrict our attention to the second task mentioned above. This task is becomes difficult by modifying the memory layout of the program by making the stack non-executable, thus preventing code injection by stack-smashing attacks. Even more better and complete defense (W$\oplus$X) ensures that no memory location in a program is marked both writable (W) and executable (X) \cite{w+x}.
Due to code injection defenses, attackers moved on to reusing code already present in the exectuable (code) region of program's memory. Thus, in principle any available code, either from the program's text segment or from a library to which it linked could be used. However, The standard C library (libc) was the usual target since it contained almost every UNIX program necessary, these attacks were thus named Return-into-libc attacks. The attacker can cause a series of functions to be invoked, one after the other by carefully setting up the stack.

Generalizing Return-into-libc attacks, Return-oriented programming technique was used to allow arbitrary (Turing complete) computation \cite{ROP-1}, without calling any functions explicitly.

\subsection{Mitigations}
We briefly discuss mitigations against memory error exploitation and their effects on Return-Oriented programming. Stack-Guard \cite{stackguard} and ProPolice \cite{propolice} are some of the techniques which prevent traditional stack-smashing attacks: preventing subversion of a program's control flow with typical buffer overflows on the stack, these defenses are good but there are known circumvention methods \cite{circum}. Bounds Checking \cite{bounds} is also a good mitigation technique which fights the problem of memory corruption at its root and tries to prevent buffer overflows. However, stack smashing is not necessary for performing a Return-oriented attack.

Address-space layout randomization (ASLR) \cite{aslr} is another relevant and widely used mitigation technique. Any attack which requires the knowledge of addresses in the program's memory space is defeated by ASLR-except for brute force search \cite{aslrbrute}, partial address overwrites \cite{aslrpartial}, and information disclosure \cite{aslrdisc}. This mitigation technique applies to both code injection and code reuse attacks equally well. Assuming effective ASLR, the presence or absence of DEP is irrelevant.

Shadow Stack \cite{shadowstack} is another mitigation technique to ensure correct execution of the program. The concept used is to have an area of memory where backup copies of all return addresses are stored, upon returning from a function, the return address stored on the stack is compared to the one on top of the shadow stack (should be identical in case of correct program execution). Using a shadow stack battles ROP attacks, because attacker would have to overwrite both the return addresses on the stack and the one on the shadow stack (typically protected). A problem which persists with this technique is that it only protects return addresses, hence does not provide protection against JOP attacks. Intel's Control-flow Enforcement Technology (CET) \cite{cet} is an example of this.

Finally, control-flow integrity \cite{XFI}, \cite{cfi-1} systems can provably prevent a program's control-flow from being hijacked. They classify indirect control-flow instructions and indirect control-flow targets into groups and enforce policies so indirect call-sites match the possible targets. This drastically reduces the number of available gadgets but still there are many attacks which can be bypassed.

\subsection{The x86\_64 Architecture}
We will introduce basics of the x86\_64 architecture \cite{intel} to aid us later in our discussion
We will cover registers, management of stack and heap and function calls. x86\_64 instructions are of variable size (depending on the instruction) which can be useful in performing ROP attacks.

\subsubsection{Registers}
Registers are used to perform faster computation, x86\_64 has 16 \textit{general purpose registers} each of which is 64 bit long, but it is possible to use lower 32 or 16 bits, lower/higher 8 bits of a register. Using lower bits only of a register also decreases the size of the instruction it is used in, such as using lower 32 bits instead of 64 bits can reduce the size of x86\_64 instruction by 2 bytes. The set of general purpose registers is as follows:

Registers \texttt{rip}, namely program counter and \texttt{rflags}, namely collection of flags are special registers in x86\_64 and are responsible for program-flow and storing information about results of previous operations respectively.

\subsubsection{Memory Management}
Memory can be allocated statically using the stack, it is very fast and handled automatically, but space is limited; or can be allocated dynamically using heap, it is slightly slower than static memory allocation, but has no limits on size, though the programmer has to manage memory.

Stack is a small area of memory used by functions to temporarily store information, local variables etc. Two registers namely \texttt{rsp} (stack pointer) and \texttt{rbp} (base or frame pointer) are used to manage stack. After each push operation, stack grows and rsp is moved down (as stack grows from higher to lower addresses) to point to the top of the stack; after each pop operation, stack shrinks and rsp is moved up to point to the top of the updated stack.

Heap does not have a fixed size, it can grow arbitrary large, but it is not managed automatically by the hardware, heap management is done in software by operating system.

\subsubsection{Function Calls}

A function call is initiated using the \texttt{call} instruction, however, it requires the program flow to return to the calling function, namely the \textit{caller}, once the called function, namely the \textit{callee}, has finished. The steps showing the working of a function call are:

\begin{enumerate}
    \setlength\itemsep{0em}
    \item Caller sets up parameters in registers and stack (if number of parameters is large) that need to be passed in accordance of the calling convention (first six arguments go in \texttt{rdi, rsi, rdx, rcx, r8, r9}, rest on stack).
    \item The \texttt{call} instruction pushes the return address on the stack and sets \texttt{rip} to the target address.
    \item Callee pushes \texttt{rbp} on the stack and creates a new stack frame for itself by setting \texttt{rbp} to \texttt{rsp}.
    \item The callee reserves some space on the stack by decreasing \texttt{rsp} and remaining instructions in the callee execute.
    \item Callee sets the stack back to the way it was by setting \texttt{rsp} to \texttt{rbp} and pops into \texttt{rbp}, restoring the original value of \texttt{rbp} which was stored earlier.
    \item The \texttt{ret} instruction loads the value \texttt{rsp} points to into \texttt{rip}, it will be the return address stored earlier, this instruction is just like \texttt{pop rip} instruction.
    \item If parameters were passed on the stack, the callee cleans up the stack also by increasing \texttt{rsp}.
\end{enumerate}

\clearpage
\section{Design}
\label{sec:design}
In this section, we will discuss the working of our tool, algorithm of how it uses the gadgets retrieved from an x86\_64 executable file to build a ROP chain of gadgets executing some arbitrary function. This process is completed in 3 steps:

\begin{itemize}
    \setlength\itemsep{0em}
    \item First step is to fetch all the ROP Gadgets from the binary upto a maximum depth.
    \item Next, process each gadget and convert it into a sequence of simpler mathematical equations and divide the gadgets into 2 categories; (1) gadgets whose result does not depend on any initial state of registers, (2) gadgets whose result depend on initial state of registers.
    \item Finally, compute a permutation of gadgets which is essentially the required ROP chain of gadgets.
\end{itemize}

\subsection{Inputs}
The tool takes 2 inputs:
\begin{itemize}
    \setlength\itemsep{0em}
    \item \textbf{Executable file}: User has to provide some x86\_64 binary file, the only restriction is that it should not reference any dynamic libraries, all the libraries required for the binary must be statically linked. The reason for static linking is that we will process the binary for gadgets, all the binary code should be available to use, if the linking is dynamic then this will not be the case.
    \item \textbf{Program}: User provides a program formatted as one or more series of system calls that he wants to execute, with arbitrary arguments, such as "exit(1)", write(1, "hello, world", 12) etc.
\end{itemize}

\subsection{Important points in algorithm}
While developing the algorithm, we have taken certain assumptions and determined some important results regarding gadgets, they are the following:
\begin{itemize}
    \item \textbf{Assumption}: Memory instructions are ignored. The reason behind this assumption is while performing static analysis of binary, one cannot deterministically find the value stored in some memory address because it is dynamic, thus it can't always be dealt with during static analysis.
    \item In all the gadgets used for analysis, the sequence of instructions should such that each \texttt{push} or stack pointer decrement instruction should have a corresponding \texttt{pop} or stack pointer increment function. If this is not followed and in case the stack has further grown during the execution of gadget, then at the time of \texttt{ret} instruction, the stack pointer won't be pointing to next gadget, thus, jumping to next gadget will not be possible since it leads to a recursive problem.
\end{itemize}

\subsection{Algorithm}
In this subsection, we will discuss the 3 steps of the algorithm in detail.

We will illustrate the algorithm using an example of \bold{echo} binary from GNU Coreutils 8.1 for ROP gadgets and use the gadgets retrieved from it to execute \bold{mprotect} system call with arbitrary arguments.

\subsubsection{Retrieval of Gadgets}
To retrieve ROP gadgets from an executable file, we use a tool called ROPGadget \cite{ROPgadget}, this tool is useful in fetching and searching for gadgets in binaries on different platforms. It retrieves all ROP Gadgets upto certain depth and gives us gadgets in x86\_64 assembly code format along with the offset address inside the binary. For illustrating the output received at this step, we will show some of the gadgets which will be used in later steps, we will assign gadget serial numbers for easy reference.

\begin{center}
    \begin{tabular}{||c|c|c||}
    \hline
        \bold{No} & \bold{Offset} & \bold{Gadget Instructions} \\ [0.5ex]
        \hline\hline
         1 & 0x00000000000054cf & \texttt{mov edx,eax; add rsp, 8; ret} \\
         \hline
         2 & 0x0000000000005011 & \texttt{mov eax, 0x1; ret} \\
         \hline
         3 & 0x00000000000026d0 & \texttt{mov eax, 0xa; ret} \\
         \hline
         4 & 0x00000000000022fe & \texttt{pop rdi; ret} \\
         \hline
         5 & 0x00000000000022fc & \texttt{pop rsi; pop r15; ret} \\
         \hline
    \end{tabular}
\end{center}

\subsubsection{Transformation of Gadgets}
The next step is to process each gadget retrieved from the binary and transform it into a sequence of simple mathematical equations to make the computation easier, these equations will be used later in this step to classify gadgets into 2 categories as described earlier in this section.

From Step 1, we will get all the gadgets in x86\_64 assembly code format, using the knowledge of x86\_64 architecture, we will convert each instruction of a gadget to a simple equation, thus each gadget will form a sequence of equations.

The conversion scheme for different x86\_64 instructions will be as follows:

\begin{itemize}
    \setlength\itemsep{0em}
    \item \bold{Data Movement Instructions}: instructions like \texttt{mov}, \texttt{push} and \texttt{pop}.
    \begin{itemize}
        \setlength\itemsep{0em}
        \item \texttt{mov} instruction \newline
            \texttt{mov dest, src} $\rightarrow$ \texttt{dest = src} \newline
            \texttt{mov dest, const} $\rightarrow$ \texttt{dest = const}
        \item \texttt{push} and \texttt{pop} instructions \newline
            We will discuss these instructions later in detail.
    \end{itemize}
    \item \bold{Arithematic \& Logic instructions}: instructions performing basic arithematic and logical operations.
    \begin{itemize}
        \setlength\itemsep{0em}
        \item \texttt{add/sub/and/or/xor} instructions \newline
            \texttt{op dest, src} $\rightarrow$ \texttt{dest = dest op src} \newline
            \texttt{op dest, const} $\rightarrow$ \texttt{dest = dest op const}
        \item \texttt{inc/dec/neg/not} instructions \newline
            \texttt{op dest} $\rightarrow$ \texttt{dest = op dest}
        \item \texttt{imul} instruction \newline
            \texttt{imul dest} $\rightarrow$ \texttt{rdx:rax = rax * dest} (rdx used in case of overflow) \newline
            \texttt{imul dest, src} $\rightarrow$ \texttt{dest = dest * src} \newline
            \texttt{imul dest, src, const} $\rightarrow$ \texttt{dest = src * const}
        \item \texttt{idiv} instruction \newline
            \texttt{idiv dest} $\rightarrow$ \texttt{rax = rdx:rax / dest}; and \texttt{rdx = rdx:rax \% dest} \newline
            The values of \texttt{rax} and \texttt{rdx} are quotient and remainder respectively.
        \item \texttt{shl/shr/sal/sar/ror/rol} instructions (shift and rotate) \newline
            \texttt{op dest, const} $\rightarrow$ \texttt{dest = dest op const} \newline
            \texttt{op dest, cl} $\rightarrow$ \texttt{dest = dest op cl} (cl is lower 8 bits of RCX register)
    \end{itemize}
    \item \bold{Exchange instructions}: instructions which exchange the values of operands.
    \begin{itemize}
        \setlength\itemsep{0em}
    \item \texttt{xchg} instruction \newline
            \texttt{xchg dest, src} $\rightarrow$ \texttt{temp = dest}; \texttt{dest = src}; \texttt{src = temp} \newline
            Basically we exchange the contents of dest and src registers.
        \item \texttt{xadd dest, src} $\rightarrow$ \texttt{temp = src + dest}; \texttt{src = dest}; \texttt{dest = temp} \newline
            In addition to exchange of values, sum is placed in dest.    
    \end{itemize}
\end{itemize}

\noindent\bold{Push and Pop Instructions} \newline
While processing a gadget, we will maintain a stack (separate for each) which will be used to imitate the stack changes during the execution of the gadget, the \texttt{push} and \texttt{pop} instructions within a gadget will be condensed into equations using this imitated stack only. An important thing to note here is that, if there is a \texttt{pop} instruction without a corresponding \texttt{push} instruction, then the value will be retrieved from stack which is under our control, thus the value can be set arbitrarily (represented by * in equation). An example of this is as follows:

\begin{center}
    \begin{tabularx}{0.8\textwidth}{||X|X||}
    \hline
        \bold{Gadgets} & \bold{Sequence of Equations} \\ [0.5ex]
        \hline\hline
         \texttt{pop rax; push rbx; pop~rdx; pop rcx; ret} & \texttt{rax = *; rdx = rbx; rcx~= *} \\
         \hline
         \texttt{push r15; push r14; pop~rax; push r13; pop~rbx; pop rcx; pop~rdx; ret} & \texttt{rax = r14; rbx = r13; rcx = r15; rdx = *} \\
         \hline
    \end{tabularx}
\end{center}

The algorithm for transformation of gadgets will be as follows:

    \begin{algorithm}[h]
      \caption{Transform Gadget}
      \begin{algorithmic}[1]
        \Procedure{transformGadget}{Gadget $\mathcal{G}$}\Comment{Takes gadget as input}
      \State s $\gets$ new Stack() \Comment{Initialize stack for gadget execution}
      \ForAll{inst $i \in \mathcal{G}.inst$}
        \If{$i$ is a "ret" inst}
            \If{s.empty $\neq$ true}
                \State $\mathcal{G}.equations \gets []$
                \State \textbf{return} false\Comment{Gadget can't be used}
            \EndIf
            \State \textbf{return} true
        \ElsIf{$i$ is a ``push reg" inst}
            \State s.push(reg)
        \ElsIf{$i$ is a ``pop reg" inst}
            \If{s.empty == true}
                \State $\mathcal{G}$.add\_equation(reg = *)\Comment{Stack is under our control}
            \Else
                \State $tmpreg \gets$ s.pop()
                \State $\mathcal{G}$.add\_equation(reg = tmpreg)
            \EndIf
        \Else
            \State $e \gets$ convert\_inst($i$)\Comment{Convert according to defined rules}
            \State $\mathcal{G}$.add\_equation($e$)
        \EndIf
      \EndFor
        \EndProcedure
      \end{algorithmic}
    \end{algorithm}
An important point to be noted here is that, during building a sequence of equations, we try to remove redundant calculations, this means that if a sequence of equations such as "\texttt{rax = 0; rbx = rax + 10;}" occurs, then we flatten the equations into "\texttt{rax = 0; rbx = 10;}".

After transformation of gadgets into simple equations, the gadgets will be classified into 2 categories, based on their dependence on previous values of registers. This classification can be performed trivially by checking if the RHS of any equation in the sequence involves a register argument.

The algorithm for classification of gadgets will be as follows:

    \begin{algorithm}[h]
      \caption{Classify Gadget}
      \begin{algorithmic}[1]
        \Procedure{classifyGadget}{Gadget $\mathcal{G}$}\Comment{Takes gadget as input}

      \ForAll{equations $e \in \mathcal{G}.equations$}
        \If{RHS has a register argument $a_r$}
            \State \textbf{return} Category 2\Comment{Dependent on initial state}
        \EndIf
      \EndFor
      \State \textbf{return} Category 1\Comment{Independent of initial state}
        \EndProcedure
      \end{algorithmic}
    \end{algorithm}

Performing the 2nd step on our example will give us following sequence of equations for each of the gadgets, simultaneously classifying gadgets into the 2 categories:

\begin{center}
    \begin{tabular}{||c|c||}
    \hline
        \bold{Gadgets} & \bold{Sequence of Equations} \\ [1ex]
        \hline\hline
        \multicolumn{2}{||c||}{\bold{Category 1 Gadgets}} \\ [0.5ex]
         \hline
         \texttt{mov eax, 0x1; ret} & \texttt{eax = 1}\\
         \hline
         \texttt{mov eax, 0xa; ret} & \texttt{eax = 10}\\
         \hline
         \texttt{pop rdi; ret} & \texttt{rdi = *}\\
         \hline
         \texttt{pop rsi; pop r15; ret} & \texttt{rsi = *; r15 = *} \\ [0.5ex]
         \hline\hline
         \multicolumn{2}{||c||}{\bold{Category 2 Gadgets}} \\ [0.5ex]
         \hline
         \texttt{mov edx,eax; add rsp, 8; ret} & \texttt{edx = eax; rsp = rsp + 8}\\
         \hline
    \end{tabular}
\end{center}

\subsubsection{Construction of ROP chain}
At this point, we have 2 classes of gadgets, one which depend on initial values of some registers and other one which do not. These gadgets are in the form of simple mathematical equations. Now, our goal is to generate a chain of gadgets which executes the program (system call) we desire.

We will denote our goal, which is execution of a system call with arbitrary arguments, as a required objective state of registers before making the syscall. For our example, the objective state will be as follows: \newline

\begin{centering}
        \texttt{rax = 10}; \texttt{rdi = addr}; \texttt{rsi = length}; \texttt{rdx = 1}
\end{centering}

\newpage We will process the first category of gadgets (having no dependence) separately, this processing step is performed mainly for 2 reasons:
\begin{itemize}
    \item We can have all the possible states of registers which do not depend on any initial state, i.e. all of these states can be generated and used as an initial state for any gadget which depends on some initial state.
    \item Processing of these gadgets is easy, and most of the binaries have such diverse gadgets that only using these gadgets can achieve the objective state, without even having to process the 2nd category of gadgets.
\end{itemize}

For each register involved in ``objective\_state", we will maintain a list of permutations which result in the required value of that register.
The algorithm for Permuting over category 1 gadgets will be as follows:

    \begin{algorithm}[h]
      \caption{Permutation1}
      \begin{algorithmic}[1]
        \Procedure{Permutation-1}{Gadget $g \in \mathcal{G}_{c1}$}\Comment{Takes Category 1 gadgets as input}
      \ForAll{permutations $p$ of gadgets $g \in \mathcal{G}_{c1}$}
        \State save\_distinct\_state($p$)\Comment{Save state only if it doesn't exist before}
      \EndFor
      \ForAll{reg $r \in \text{objective\_state.registers}$}
          \ForAll{distinct permutations $p$ including $r$ (from saved states)}
            \If{$p.\text{state.r.val} == \text{r.val} $}
            \State r.permutations.append($p$)
            \EndIf
          \EndFor
      \EndFor
        \EndProcedure
      \end{algorithmic}
    \end{algorithm}

After processing the first category of gadgets, we have the set of states which can be used as an initial state at any point of time for the 2nd category of gadgets. We will now process 2nd category of gadgets only for those registers whose values are not set from the 1st category of gadgets.

For each register whose value is not reached the final state by a permutation of gadgets from category 1, we will start picking gadgets from category 2, each distinct permutation of category 2 will be evaluated on the basis of what initial state will result in a required final state, if the initial state is achievable, then save the current permutation for the register. During this permutation, we will form a sequential structure from the permutation of category 2 gadgets whose root will be finally a category 1 gadget which will signify the initial state that when combined with remaining elements (category 2 gadgets) produces the required result.
The algorithm for permuting over category 2 gadgets will be as follows:

    \begin{algorithm}[h]
      \caption{Permutation2}
      \begin{algorithmic}[1]
        \Procedure{Permutation-2}{Gadget $g_1 \in \mathcal{G}_{c1}, g_2 \in \mathcal{G}_{c2} $}
      \ForAll{Register  $r \in \text{final\_states.registers}$ (if still not set)}
                \State Permute over Category 2 Gadgets involving $r$ in LHS and find out what initial state does each permutation require to reach final\_state.
                \If{initial state is achievable}
                    \State Save the current permutation for the current register.
                \EndIf
      \EndFor
        \EndProcedure
      \end{algorithmic}
    \end{algorithm}

After we have achieved the final state required for making the function call or system call (in our case), we will set the stack as required by the program and output the generated final stack to the user.

The stack layout for our example will be as follows:

\begin{center}
    \begin{tabular}{||c||}
    \hline
         \texttt{Gadget \# 3}\\
         \hline
         dummy value (rsp+8)\\
         \hline
         \texttt{Gadget \# 1}\\
         \hline
         \texttt{Gadget \# 2}\\
         \hline
         dummy value (r15)\\
         \hline
         \texttt{length (= rsi)} \\
         \hline
         \texttt{Gadget \# 5} \\
         \hline
         \texttt{addr (= rdi)} \\
         \hline
         \texttt{Gadget \# 4}\\
         \hline
    \end{tabular}
\end{center}

First gadget to be executed in the above stack will be \texttt{Gadget \# 4} since stack grows downwards.
At the top of the stack we can make a call to the gadget which performs a syscall.

\clearpage
\section{Evaluation}
\label{sec:evaluation}

For evaluating our tool, we have tested our tool against 35 binaries from GNU Coreutils 8.1.
For testing of all the binaries, we kept the maximum depth of ROP Gadget as 10 bytes, if we increase the depth we will get more gadgets having no effect on current gadgets.
The exhaustive list of binaries with the total number of ROP gadgets retrieved is shown in the graph below.

From the retrieved list of gadgets, the number of gadgets used for analysis, i.e. gadgets which were transformed into sequence of equations to be processed later are is shown in the graph below.

\begin{figure}[h]
    \includegraphics[scale=0.6]{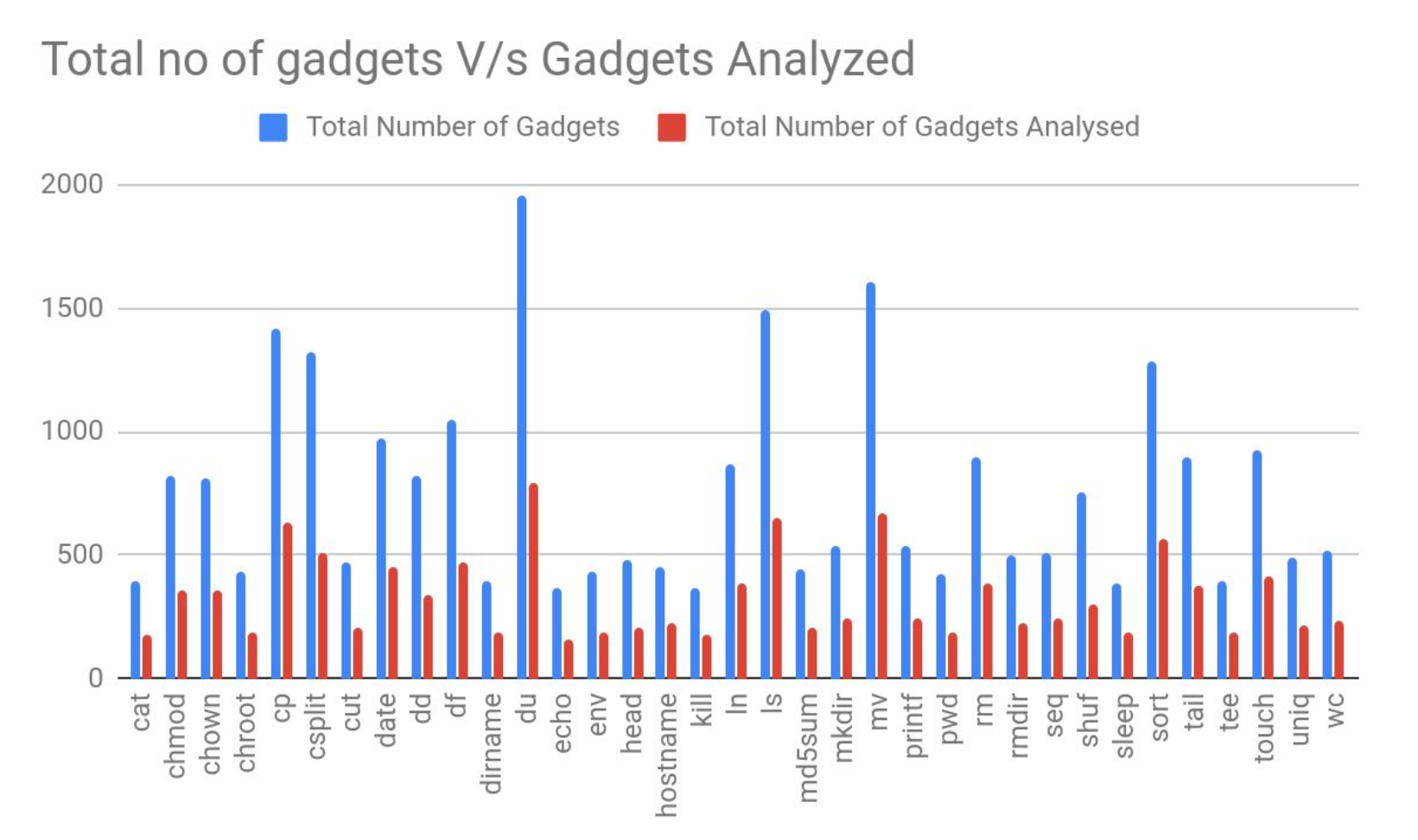}
    \centering
    \caption{Total number of gadgets fetched from the binary}
\end{figure}

The numbers of category 1 and category 2 in binaries are shown in graph below.

\begin{figure}[h]
    \includegraphics[scale=0.6]{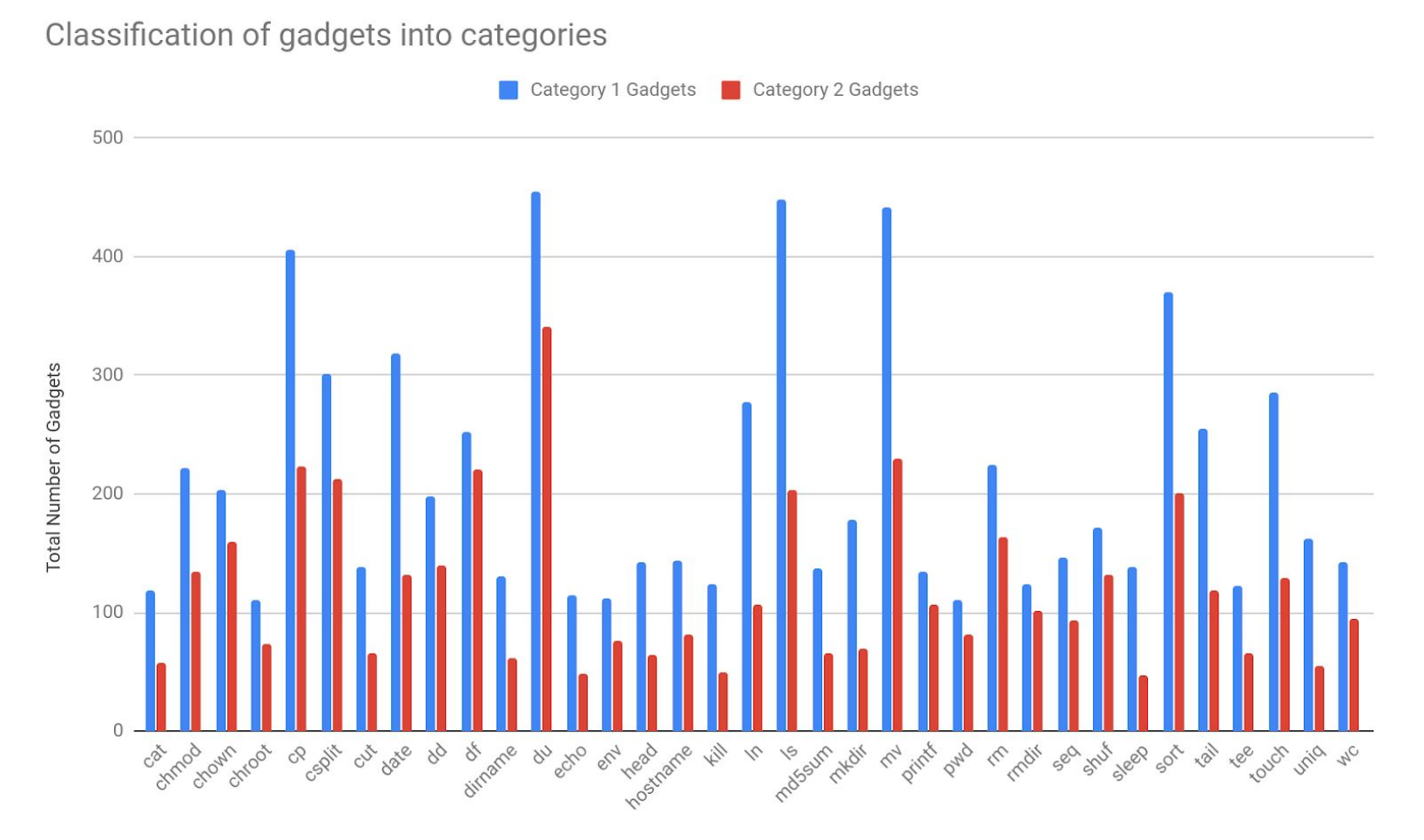}
    \centering
    \caption{Total number of gadgets fetched from the binary}
\end{figure}

For each of the programs the length of ROP chain generated for executing the same syscall \textit{mprotect} remains almost same.
The reason behind above result is that each binary has a high amount of pop and mov instructions which sets the final state for most of the required registers.

\clearpage
\section{Related Work}
\label{sec:related}

As we discussed earlier, the technique of return-to-libc attacks does not rely on code injection, it redirects the control flow of the program to one or more functions found in the standard library (standard C library in case of Linux), allowing the attacker to compromise the target program. Later research \cite{rtlibc} has also shown that these sets of gadgets are Turing Complete when the target binaries are using the standard C library.

The first introduced return-to-libc attacks used entire library functions as the basic elements of the attack code (i.e. a gadget consisted of the entire library function, not just a fragment). Krahmer \cite{x64buffer} uses small snippets of code from inside libraries or the program itself, instead of using the entire functions, these snippets were called borrowed code chunks, each chunk was responsible for transferring the execution to the next chunk. Shacham \cite{functions} formalizes these attacks and shows that Turing-Completeness is possible even if the C library of Linux system is used, he created chunks ending in a \texttt{ret} instruction, our so-called ROP gadgets. There was also some follow up work in RISC architectures \cite{risc-1}, \cite{risc-2}, \cite{ROP-3}, \texttt{pop/jmp} paired instructions and Jump-oriented programming \cite{JOP-1}.

The first ROP paper \cite{functions} describes a hand-picked set of gadgets. The arrangement or presence of certain gadgets may depend on compilers or compilation options. Shacham, in the original paper uses combinations of complex gadgets to implement arithmetic operations and branching.

There were more efforts focussed on automated construction of valid and Turing-complete sets of gadgets \cite{few}, \cite{red}, \cite{seman}, \cite{smt}. These tools usually worked with small programs written in some specific format, taking gadgets out of an input binary and generating a suitable chain of gadgets implementing the program.

Most of the approaches faced a significant problem of matching gadgets to execute some arbitrary program, most of them limit the gadgets to a few instructions \cite{few}
. They only use their limited set of gadgets for each required operations. Another implementation \cite{red} reduces each gadget to an expression tree which captures the gadget's behaviour. Another one \cite{seman} identifies the semantics of a gadget by using a semantic analysis on postconditions, then a similar tree-based approach to match the semantics to the behavior of the code. An other approach \cite{smt} uses a SMT solver to re-write each gadget as a series of binary functions of the input bits, each function corresponds to an output bit, thus, the problem reduces to matching boolean functions between the gadgets and the target code.

\clearpage
\section{Conclusion/Future Work}
\label{sec:conclusion}

We presented our tool which takes as input a binary and a program in the form of system calls, gets a list of ROP gadgets present in the binary upto a certain maximum depth (can be set arbitrarily) and uses these gadgets to build a ROP chain of gadgets which can be used to execute the program using already existing code in the program's memory space. The gadgets were converted to a set of equations which can be evaluated mathematically and using these equations, it can be found which gadgets depend on some initial state of registers and which ones do not. Thus, using this info we figure out a suitable permutation of gadgets which will satisfy a final state for executing the code that we desire.

This tool was designed based on the principles of ROP and x86\_64 architecture, its purpose is to tackle the problems that a user faces when developing a program using ROP and in turn automate this process so that user does not have to devote a lot of his/her time in finding gadgets and placing them in suitable order. The purpose of this tool is also to verify that ROP gadgets are Turing Complete, for which we take an arbitrary binary from user and try to find gadgets only from that by static linking libraries (if any used).

Our approach is not a very efficient one, since the time taken in finding suitable ROP chain of gadgets for a large binary can take a long time. We plan to make the tool more efficient by introducing caching techniques and extending the tool in multiple dimensions i.e. the computation involving gadgets which do not depend on each other should be done in parallel, not in sequential manner.

Our design currently assumes various mitigation techniques such as ASLR, shadow stack and others are disabled and our focus is only on finding a suitable ROP chain of gadgets which executes a given program by reusing the code available in the binary. In future, we can expand the design of the tool to include attacks against commonly used mitigation techniques, so even if any of the mitigation techniques are present, a suitable ROP attack can still be performed.

We currently don't support building chains using Jump-Oriented programming technique but we can retrieve JOP gadgets, in future, we would integrate JOP, so that the chain of gadgets generated might take advantage of both JOP and ROP, because now-a-days there are mitigation techniques which prevent ROP attacks but they do not work well with JOP attacks.

In future, we plan to test our tool against many more benchmarks to improve its quality and thus fulfill our goals of finding difficulties and fixing them by completing an automated tool which can produce a suitable ROP chain of gadgets for any program with any binary as input, thus verifying that ROP is indeed Turing Complete.



\clearpage
\renewcommand*{\thesection}{}\textbf{}

\bibliographystyle{alpha}
\bibliography{Bibliography}

\begin{thebibliography}{JWM12}

\bibitem[Ano01]{Heap-2}
Anonymous.
\newblock {Once upon a free(). . . .}
\newblock {\em Phrack Magazine 57}, 2001.

\bibitem[BK00]{circum}
Bulba and Kil3r.
\newblock {Bypassing StackGuard and StackShield}.
\newblock {\em Phrack Magazine 56}, 2000.

\bibitem[Bla10]{aslrdisc}
Dionysus Blazakis.
\newblock {Interpreter exploitation}.
\newblock {\em Proceedings of WOOT 2010, H. Shacham and C. Miller, Eds.
  USENIX}, 2010.

\bibitem[Ble02]{int-3}
Blexim.
\newblock {Basic integer overflows}.
\newblock {\em Phrack Magazine 60}, 2002.

\bibitem[CC98]{stackguard}
Dave Maier Jonathan Walpole Peat Bakke SteveBeattie Aaron Grier Perry Wagle
  Qian~Zhang Crispan~Cowan, Calton~Pu.
\newblock {StackGuard: Automatic detection and prevention of buffer-overflow
  attacks}.
\newblock {\em Proceedings of USENIX Security 1998}, pages 63--78, 1998.

\bibitem[Des00]{Heap-1}
Solar Designer.
\newblock Jpeg com marker processing vulnerability in netscape browsers.
\newblock
  \url{https://www.openwall.com/articles/JPEG-COM-Marker-Vulnerability}, 2000.

\bibitem[Dur02]{aslrpartial}
Tyler Durden.
\newblock {Bypassing PaX ASLR protection}.
\newblock {\em Phrack Magazine 59}, 2002.

\bibitem[EB08]{risc-1}
Hovav Shacham Stefan~Savage Erik~Buchanan, Ryan~Roemer.
\newblock { When good instructions go bad: generalizing return-oriented
  programming to RISC}.
\newblock {\em Proceedings of the 15th ACM Conference on Computer and
  Communications Security}, pages 27--38, 2008.

\bibitem[EJS11]{seman}
David~Brumley Edward J.~Schwartz, Thanassis~Avgerinos.
\newblock {Q: Exploit Hardening Made Easy}.
\newblock {\em Proceedings of the 20th USENIX Security Symposium (2011), USENIX
  Association}, 2011.

\bibitem[GR01]{str-2}
Gera and Riq.
\newblock {Advances in format string exploiting.}
\newblock {\em Phrack Magazine 59}, 2001.

\bibitem[HE01]{propolice}
Kunikazu~Yoda Hiroaki~Etoh.
\newblock {ProPolice: Improved stack-smashing attack detection}.
\newblock 2001.

\bibitem[Hor02]{int-2}
Oded Horovitz.
\newblock {Big loop integer protection}.
\newblock {\em Phrack Magazine 60}, 2002.

\bibitem[HS04]{aslrbrute}
Ben Pfaff Eu-Jin Goh Nagendra Modadugu Dan~Boneh Hovav~Shacham, Matthew~Page.
\newblock {On the effectiveness of address-space randomization}.
\newblock {\em Proceedings of CCS 2004}, page 298–307, 2004.

\bibitem[Int16a]{cet}
Intel.
\newblock {Control-flow enforcement technology preview}.
\newblock 2016.

\bibitem[Int16b]{intel}
Intel.
\newblock {Intel 64 and IA-32 Architectures Software Developer's Manual}.
\newblock 2016.

\bibitem[JWM12]{JOP-3}
Dong-Young Lee Tai-Myoung~Chung Jae-Won~Min, Sung-Min~Jung.
\newblock {Jump oriented programming on windows platform (on the x86)}.
\newblock {\em Lecture Notes in Computer Science}, 7335:376–390, 2012.

\bibitem[Kae01]{Heap-3}
Michel Kaempf.
\newblock {Vudo malloc tricks}.
\newblock {\em Phrack Magazine 57}, 2001.

\bibitem[Klo99]{frame}
Klog.
\newblock {The frame pointer overwrite}.
\newblock {\em Phrack Magazine 55}, 1999.

\bibitem[Kor09]{ROP-3}
Tim Kornau.
\newblock {Return oriented programming for the arm architecture.}
\newblock {\em Master’s thesis, RuhrUniversität Bochum}, 2009.

\bibitem[Kra05]{x64buffer}
Sebastian Krahmer.
\newblock {x86-64 buffer overflow exploits and the borrowed code chunks
  exploitation technique}.
\newblock {\em Online}, 2005.

\bibitem[LD10]{risc-2}
Ahmad-reza Sadeghi-Marcel~Winandy Lucas~Davi, Ra~Dmitrienko.
\newblock {Return-oriented programming without returns on ARM}.
\newblock {\em System Security Lab, Ruhr University Bochum, Germany}, 2010.

\bibitem[MA09]{cfi-1}
Úlfar Erlingsson Jay~Ligatti Martín~Abadi, Mihai~Budiu.
\newblock {Control-flow integrity principles, implementations, and
  applications}.
\newblock {\em ACM Trans. Info. \& System Security 13}, 2009.

\bibitem[MT11]{rtlibc}
Tyler Bletsch Xuxian Jiang Vincent Freeh-Peng~Ning Minh~Tran, Mark~Etheridge.
\newblock {On the Expressiveness of Return-into-libc Attacks}.
\newblock {\em Proceedings of the 14th International Symposium on Recent
  Advances in Intrusion Detection}, 2011.

\bibitem[One96]{Aleph}
Aleph One.
\newblock {Smashing the stack for fun and profit}.
\newblock {\em Phrack, 7}, 1996.

\bibitem[PA09]{bounds}
Miguel Castro Steven~Hand Periklis~Akritidis, Manuel~Costa.
\newblock {Baggy bounds checking: An efficient and backwards-compatible defense
  against out-of-bounds errors}.
\newblock {\em 18th USENIX Security Symposium, Montreal, Canada, August 10-14,
  2009, Proceedings}, pages 51--66, 2009.

\bibitem[RH09]{few}
Felix C.~Freiling Ralf~Hund, Thorsten~Holz.
\newblock {Return-Oriented Rootkits: Bypassing Kernel Code Integrity Protection
  Mechanisms}.
\newblock {\em Proceedings of the 18th USENIX Security Symposium}, pages
  383--398, 2009.

\bibitem[RR12]{ROP-1}
Hovav Shacham Stefan~Savage Ryan~Roemer, Erik~Buchanan.
\newblock {Return-oriented programming: Systems, languages, and applications}.
\newblock {\em ACM Trans. Inf. Syst. Secur.}, 2012.

\bibitem[SA04]{DEP}
Vincent~Abella Starr~Andersen.
\newblock {Changes to Functionality in Microsoft Windows XP Service Pack 2}.
\newblock page Part 3, April, 2004.

\bibitem[Sal11]{ROPgadget}
Jonathan Salwan.
\newblock Ropgadget.
\newblock \url{https://github.com/JonathanSalwan/ROPgadget}, 2011.

\bibitem[SC10]{JOP-2}
Alexandra Dmitrienko Ahmad-Reza Sadeghi Hovav Shacham Marcel~Winandy
  Stephen~Checkoway, Lucas~Davi.
\newblock {Return-oriented programming without returns}.
\newblock {\em Proceedings of the 17th ACM Conference on Computer and
  communications security, CCS ’10}, page 559–572, 2010.

\bibitem[Sha07a]{ROP-2}
Hovav Shacham.
\newblock {m. The geometry of innocent flesh on the bone: return-into-libc
  without function calls (on the x86)}.
\newblock {\em Proceedings of the 14th ACM Conference on Computer and
  communications security, CCS ’07}, page 552–561, 2007.

\bibitem[Sha07b]{functions}
Hovav Shacham.
\newblock {The Geometry of Innocent Flesh on the Bone: Return-into-libc without
  Function Calls (on the x86)}.
\newblock {\em Proceedings of the 14th ACM Conference on Computer and
  Communications Security}, pages 552--561, 2007.

\bibitem[Sol10]{smt}
Pablo Sole.
\newblock {Hanging on a ROPe}.
\newblock {\em ekoParty Security Conference}, 2010.

\bibitem[ST01]{str-1}
Scut and Team Teso.
\newblock {Exploiting format string vulnerabilities}.
\newblock 2001.

\bibitem[TB11]{JOP-1}
Vince W. Freeh-Zhenkai~Liang Tyler~Bletsch, Xuxian~Jiang.
\newblock {Jump-oriented programming: a new class of code-reuse attack}.
\newblock {\em Proceedings of the 6th ACM Symposium on Information, Computer
  and Communications Security}, pages 30--40, 2011.

\bibitem[TD10]{red}
Ralf-Philipp~Weinmann Thomas~Dullien, Tim~Kornau.
\newblock {A framework for automated architecture-independent gadget search}.
\newblock {\em Proceedings of the 4th USENIX conference on Offensive
  technologies}, 2010.

\bibitem[Tea03a]{aslr}
PaX Team.
\newblock {PaX address space layout randomization}.
\newblock 2003.

\bibitem[Tea03b]{w+x}
PaX Team.
\newblock {PaX non-executable pages design \& implementation}.
\newblock 2003.

\bibitem[THD15]{shadowstack}
David~Wagner Thurston H.Y.~Dang, Petros~Maniatis.
\newblock {The performance cost of shadow stacks and stack canaries}.
\newblock {\em Proceedings of the 10th ACM Symposium on Information, Computer
  and Communications Security, ASIA CCS ’15}, page 555–566, 2015.

\bibitem[UE06]{XFI}
Michael Vrable Mihai Budiu George C.~Necula Ulfar~Erlingsson, Martin~Abadi.
\newblock {XFI: Software Guards for System Address Spaces}.
\newblock {\em Proceedings of OSDI 2006}, pages 75--88, 2006.

\bibitem[Zal01]{int-1}
Michal Zalewski.
\newblock {Remote vulnerability in SSH daemon CRC32 compression attack
  detector}.
\newblock 2001.

\end{thebibliography}

\end{document}